\documentclass[aps,pra,11pt]{revtex4-2}
\usepackage{graphicx}
\usepackage{color}
\usepackage{mathrsfs}
\usepackage{amsmath}
\usepackage{amsfonts,amssymb}
\usepackage{bm}
\begin{document}
\title{Phase-space complexity of discrete-variable quantum states and operations}
\author{Siting Tang$^{1,2,3}$, Shunlong Luo$^{1,2}$, and Matteo G. A. Paris$^{3,*}$}
\affiliation{$^{1}$State Key Laboratory of Mathematical Sciences, Academy of Mathematics and Systems Science, Chinese Academy of Sciences,  Beijing 100190, China}
\affiliation{$^{2}$School of Mathematical Sciences, University of Chinese Academy of Sciences, Beijing 100049, China}
\affiliation{$^{3}$Dipartimento di Fisica, Universit\`a  di Milano, I-20133 Milano, Italy}
\email{matteo.paris@fisica.unimi.it}

\begin{abstract}
We introduce a quantifier of phase-space complexity for discrete-variable (DV) quantum systems. Motivated by a recent framework developed for continuous-variable systems, we construct a complexity measure of quantum states based on the Husimi $Q$-function defined over spin coherent states. The quantifier combines 
into a single scalar quantity two complementary information-theoretic quantities, the Wehrl entropy, which captures phase-space spread, and the Fisher information, which captures localization.  We derive  fundamental properties of this measure, including its invariance under SU(2) displacements. The 
complexity is normalized such that coherent states have unit complexity, while the completely mixed state has zero complexity, a feature distinct from the continuous-variable case. We provide analytic expressions for several relevant families of states, including Gibbs and Dicke states, and perform a numerical analysis of spin-squeezed states, NOON states, and randomly generated states. Numerical results reveal a monotonic, but not deterministic, relationship between complexity and purity, leading us to conjecture that maximal complexity is attained by pure states, thereby connecting the problem to the optimization of Wehrl entropy via Majorana constellations. Finally, we extend the framework to quantum channels, defining measures for both the generation and breaking of complexity. We analyze the performance of common unitary gates and the amplitude damping channel, showing that while low-dimensional systems can achieve maximal complexity via spin squeezing or NOON states, this becomes impossible in higher dimensions. These results highlight dimension-dependent limitations in the generation of phase-space complexity and establish a unified phase-space approach to complexity across both continuous and discrete variable regimes.
\end{abstract}

\maketitle

\section{Introduction}

The notion of statistical complexity was originally introduced to characterize physical systems 
that lie between perfect order and complete disorder \cite{lopez1995statistical,catalan2002features,romera2004fisher,manzano2012statistical}. The 
definition is made by formulating the opposite end: what is simple? The answer is, both perfect 
crystal and ideal gas are simple, following the idea that the former represents a completely 
ordered system and the latter corresponds to a completely disordered system \cite{lopez1995statistical}. 
This concept captures a fundamental trade-off: a simple system is either perfectly localized or perfectly spread, while statistical complexity emerges from the coexistence of localization and delocalization.

In quantum mechanics, phase-space distributions offer a natural setting for such a characterization. The Husimi $Q$-function \cite{husimi1940some}, in particular, is a nonnegative and informationally complete representation of a quantum state, making it suitable for defining complexity measures. The idea stems from the fact that the phase-space distributions are faithful representations of the quantum states, and among the family of phase-space distributions the Husimi $Q$-distribution is of particular interest due to its nonnegativity \cite{husimi1940some,cahill1969density}.
In a previous work \cite{Tang_2025}, this framework has been applied to continuous-variable (CV) systems, combining the Wehrl entropy \cite{wehrl1979relation} (differential entropy of the Husimi function) and the Fisher information \cite{fisher1925theory} with respect to phase-space displacements. The resulting measure quantifies how a state balances spread and localization, revealing features distinct from standard nonclassicality or non-Gaussianity measures.

Discrete-variable systems, however, have remained outside this paradigm, despite their importance in quantum computation, metrology, and foundational studies. So naturally one wants to make a similar attempt in the discrete-variable (DV) system. The key to extending the phase-space approach to DV systems lies in the use of spin coherent states, also referred to as SU(2) coherent states, atomic coherent states and Bloch coherent states \cite{radcliffe1971some,arecchi1972atomic,perelomov1972coherent,zhang1990coherent}, which represent the natural DV analogs of canonical coherent states. These states form an overcomplete basis for spin-$j$ systems and inherit a clear geometric structure via the Bloch sphere. This allows for a Husimi-like $Q$-function, Wehrl entropy, and Fisher information to be defined in a fully discrete setting. Then everything follows naturally and one can define a quantifier for the phase-space complexity in DV systems.

In this paper, we develop a self-contained theory of phase-space complexity for DV systems. We define a complexity quantifier using the Wehrl entropy and Fisher information derived from the spin Husimi function. We establish its basic properties, including invariance under SU(2) displacements, and compute it explicitly for relevant families of states: qubits, Dicke states, thermal states, spin-squeezed states, and NOON states. We explore the dependence of complexity on purity and dimension, and conjecture that the maximum complexity is achieved by pure states, thereby relating the problem to the optimization of Wehrl entropy over Majorana constellations. Our findings reveal fundamental differences between DV and CV complexity, particularly regarding the existence of minimal and maximal complexity states, and identify dimension-dependent limitations of common quantum resources.

The remainder of the work is arranged as follows. In Section II, we briefly review the formalism of spin coherent states and the associated Husimi $Q$-function for SU(2) systems. We then introduce the phase-space complexity quantifier, defined as the product of an entropy power derived from the Wehrl entropy and a normalized Fisher information. We establish its basic properties, including its invariance under SU(2) displacements and its normalization such that coherent states yield unit complexity. In Section III, we provide a characterization of the complexity across a range of DV states. We begin with a full analytic treatment of the qubit system, where complexity is shown to increase monotonically with purity. We then compute the complexity analytically for Dicke and Gibbs equilibrium states, and numerically for spin-squeezed states, NOON states, and large ensembles of randomly generated states. Based on these results, we conjecture that maximal complexity is attained by pure states, linking the problem to the long-standing open question of maximizing Wehrl entropy via optimal Majorana constellations on the Bloch sphere. We further demonstrate that common resources for quantum metrology, such as spin squeezing and NOON states, are capable of saturating this maximal complexity only in low-dimensional systems. In Section IV, we extend the framework to quantum channels, introducing separate measures for the ability of a channel to generate and to destroy complexity. We evaluate these measures for several unitary gates—including the Pauli gates, Fourier transform, phase gate, and squeezing operators, as well as for the amplitude damping channel. Our analysis shows that while unitary gates cannot destroy complexity, the amplitude damping channel exhibits a nontrivial trade-off: in higher dimensions, it can not only reduce but also, counterintuitively, generate complexity from coherent inputs, a phenomenon not captured by purity alone. Finally, in Section V, we discuss the conceptual differences between DV and CV complexity, the physical interpretation of the optimal constellations, and the dimension-dependent limitations of common quantum resources to generate maximal complexity. We conclude with open questions regarding the tight upper bound on complexity and its relation to other measures of quantumness.

\section{Complexity of DV states}
We start with a brief review of spin coherent states.
Given the group
\begin{equation*}
	G={\rm SU(2)}=\Big\{g=
		\begin{pmatrix}
			\alpha & \beta \\
			-\bar{\beta} & \bar{\alpha}
		\end{pmatrix} : \alpha,\beta \in \mathbb{C}, |\alpha|^2+|\beta|^2=1 \Big\},
\end{equation*}
for a fixed $j \in \frac12 \mathbb{N}$, we consider the unitary irreducible representation  $(T^j,\mathcal{H}^j)$ of the group $G$, with ${\rm dim} T^j = 2j+1$ \cite{hall2015lie}. The canonical basis of the representation space $\mathcal{H}^j$ is denoted by $\{ |j,\mu \rangle, \mu=-j,-j+1,\cdots,j \}.$ They are also called the Dicke states. The infinitesimal rotation operators $J_1,J_2,J_3$ satisfy the commutation relation
\begin{equation*}
	[J_l,J_m] = i \sum_{n=1}^3 \epsilon_{lmn} J_n,
\end{equation*}
where $\epsilon_{lmn}$ is the Levi-Civit\`a symbol (antisymmetric symbol). The Dicke states are eigenvectors for the operator $J_3$ and $\boldsymbol{J}^2=J_1^2+J_2^2+J_3^2$,
\begin{equation*}
	J_3 |j,\mu\rangle = \mu |j,\mu\rangle, \qquad \boldsymbol{J}^2 |j,\mu\rangle = j(j+1) |j,\mu\rangle,
\end{equation*}
and they are transformed by the raising and lowering operators $J_\pm = J_1 \pm i J_2$ via
\begin{equation*}
	J_+ |j,\mu\rangle =\sqrt{(j-\mu)(j+\mu+1)}  |j,\mu+1\rangle, \qquad J_- |j,\mu\rangle = \sqrt{(j+\mu)(j-\mu+1)}|j,\mu-1\rangle.
\end{equation*}

The orbit of $|j,-j\rangle$ under the action of the group, i.e., $\{T^j(G)|j,-j\rangle\}$, constitutes a coherent state system \cite{Perelomov1977}. Ignoring the global phase, the quotient space ${\rm SU(2)}/{\rm U(1)}$ is the set
\begin{equation*}
	\Big\{ g_\Omega =
	\begin{pmatrix}
		\cos \frac{\theta}{2} & \sin \frac{\theta}{2} e^{i \phi} \\
		-\sin \frac{\theta}{2} e^{-i \phi} & \cos \frac{\theta}{2}
	\end{pmatrix} : \theta \in [0,\pi), \phi \in [0,2\pi) \Big\},
\end{equation*}
which is isomorphic to the Bloch sphere $S^2$. Therefore a coherent state can be represented as a point $\Omega=(\sin \theta \cos \phi, \sin \theta \sin \phi, \cos \theta) \in S^2$. Let ${\rm d} \Omega = \sin \theta {\rm d} \theta {\rm d} \phi$ denote the measure on $S^2$.
The coherent states can be expanded in the canonical basis
\begin{equation}
    \label{eq:CSexpandDicke}
    |\Omega\rangle = \sum_{\mu=-j}^j 
    \sqrt{{{2j} \choose {j-\mu}}} \left( \cos \frac{\theta}{2} \right)^{j-\mu} \left( \sin \frac{\theta}{2} \right)^{j+\mu} e^{i(j+\mu)\phi} |j,\mu\rangle,
\end{equation}
and they satisfy the following resolution of unity
\begin{equation}
	\label{eq:resolution_unity}
    \int _{S^2}|\Omega\rangle\langle\Omega| \frac{2j+1}{4\pi} {\rm d}\Omega = \mathbf{1}_{2j+1},
\end{equation}
where $\mathbf{1}_{2j+1}$ denotes the identity operator on $\mathcal{H}^j$.

Next, we give the ingredients of the phase-space complexity of a quantum state $\rho$. The phase-space distribution we use here is the Husimi $Q$-function \cite{husimi1940some,cahill1969density}, which is defined based on the coherent states,
\begin{equation}
    Q(\Omega | \rho) = \langle \Omega| \rho | \Omega \rangle,
\end{equation}
and by the resolution of unity Eq. (\ref{eq:resolution_unity}), we easily see it is normalized as
\begin{equation*}
	\int  _{S^2} Q(\Omega | \rho) \frac{2j+1}{4\pi} {\rm d}\Omega = 1.
\end{equation*}

The Wehrl entropy \cite{wehrl1979relation} is defined as the differential entropy of $ Q(\Omega | \rho)$,
\begin{equation}
    S_{\rm W} (\rho) = - \int _{S^2}Q(\Omega | \rho) \ln Q(\Omega | \rho) \frac{2j+1}{4\pi} {\rm d}\Omega.
\end{equation}

The range of the Wehrl entropy is
\begin{equation}
	\frac{2j}{2j+1} \leq S_{\rm W} (\rho) \leq \ln (2j+1).
\end{equation}
The lower bound is achieved by \cite{lieb2014proof} and only by \cite{frank2023sharp,KULIKOV2025110423} coherent states, while the upper bound is achieved by the completely mixed state $\rho_{\rm m}=\frac{1}{2j+1}\mathbf{1}_{2j+1}$, due to the concavity of the integrand.

Then, we define the Fisher information based on the Husimi $Q$-function \cite{fisher1925theory}. To be more specific, it is the trace of the Fisher information matrix of the location parameters $\Omega \in S^2$, and in this case, the Fisher information is in fact independent of the parameters \cite{frieden1998physics}, hence
\begin{equation}
    \label{eq:def_Fisher}
	I(\rho) = \int  _{S^2}\frac{||\nabla Q(\Omega | \rho)||^2}{Q(\Omega | \rho)} \frac{2j+1}{4\pi} {\rm d}\Omega,
\end{equation}
where $\nabla Q = \left(\partial_\theta Q,\frac{1}{\sin\theta}\partial_\phi Q\right)$ is the gradient operator on the unit sphere, and $||\cdot||$ denotes the Euclidean norm.
We prove in Appendix A that for any pure state $|\psi\rangle \in \mathcal{H}^j$, the Fisher information is a constant, $I(|\psi\rangle \langle \psi|)=2j$. On the other hand, since the Fisher information is strictly convex \cite{cohen2006fisher}, the minimal value is attained by the completely mixed state $\rho_{\rm m}$. Therefore,
\begin{equation}
	0 \leq I (\rho) \leq 2j.
\end{equation}

We are now ready to introduce a quantifier of phase-space complexity.

{\bfseries Definition 1.} The complexity of a quantum state $\rho$ is defined as
\begin{equation}
	\mathcal{C} (\rho) = e^{S_{\rm W} (\rho) - \frac{2j}{2j+1}}  \frac{I(\rho)}{2j}.
\end{equation}

Throughout the paper, we will use the shorthand $\mathcal{C} (\psi)$ instead of $\mathcal{C} (|\psi\rangle\langle\psi|)$ for pure states $|\psi\rangle$. From the above definition it is immediate that the complexity of a coherent state is normalized at 1,
\begin{equation*}
	\mathcal{C} (\Omega) = 1, \qquad \forall \ \Omega \in S^2.
\end{equation*}
But unlike in the CV system \cite{Tang_2025}, here the coherent states are no longer the least complex states. In fact, the minimal value is given by the completely mixed state $\rho_{\rm m}$, 
\begin{equation*}
	\mathcal{C} (\rho_{\rm m}) = 0,
\end{equation*}
whose Husimi $Q$-function is the uniform distribution on the sphere, i.e., $Q(\Omega | \rho_{\rm m}) = \frac{1}{2j+1}.$

The definition of the phase-space complexity is based on the Husimi $Q$-function, and the $Q$-function can obtained via a projection on coherent states. In DV systems, this role is naturally played by the spin coherent states, which are generated by the unitary irreducible representations of the Lie group SU(2). 

To emphasize the role of coherent states as the ``unit'' we have normalized the complexity of coherent states to be 1. This suggests that a value in $[0,1)$ gives us a ``simple'' state. However in the CV system this is not possible due to the isoperimetric inequality, so coherent states are indeed the simplest states there. On the other hand, in DV systems we have found such simple states, and the extreme case is the completely mixed state, with no complexity at all. Intuitively, these states are simpler since we are introducing more classicality in them by a more uniform way of mixing, while in CV systems this cannot be done because the Hilbert space is infinite-dimensional and a ``too uniform'' probability distribution cannot be normalized properly. Mathematically, the compactness of SU(2) guarantees that all its irreducible representations are finite-dimensional \cite{hall2015lie}.

In Ref. \cite{goldberg2020extremal}, the Wehrl entropy serves as an indicator of quantumness, thus the (pure) state with the optimal constellation is regarded as the most quantum state. Troubles might arise if we include mixed states into our consideration, because the Wehrl entropy is concave. Of course we do not want the completely mixed state, which has the largest Wehrl entropy, to be the most quantum state. So it is necessary to introduce another factor as a trade-off. In this work, we have used the Fisher information. It has the desirable feature of being constant for all pure states and strictly convex under mixing. Therefore our complexity quantifier can be viewed as a measure of quantumness that applies to all states, and it coincides (up to a one-to-one exponential map) with the initial Wehrl entropy on the set of all pure states.

The complexity quantifier has the following invariance property.

{\bfseries Proposition 1.} The complexity ${\cal C}(\rho )$ is invariant under displacements,
\begin{equation}
	\mathcal{C} (D_\Omega \rho D_\Omega^\dag) = \mathcal{C} (\rho),
\end{equation}
where $D_\Omega=T^j (g_\Omega)=e^{\frac12\theta e^{i\phi}J_- - \frac12 \theta e^{-i\phi}J_+}, \Omega \in S^2$ are the SU(2) displacement operators \cite{Perelomov1977}.
The proof is straightforward by noticing that the displacement operators act as rotations on the Bloch sphere, and the measure ${\rm d}\Omega$ is invariant under rotations.

\section{Illustrative Examples}

\subsection{Qubit}
\begin{figure}[!h]
	\centering
	\includegraphics[width=0.5\textwidth]{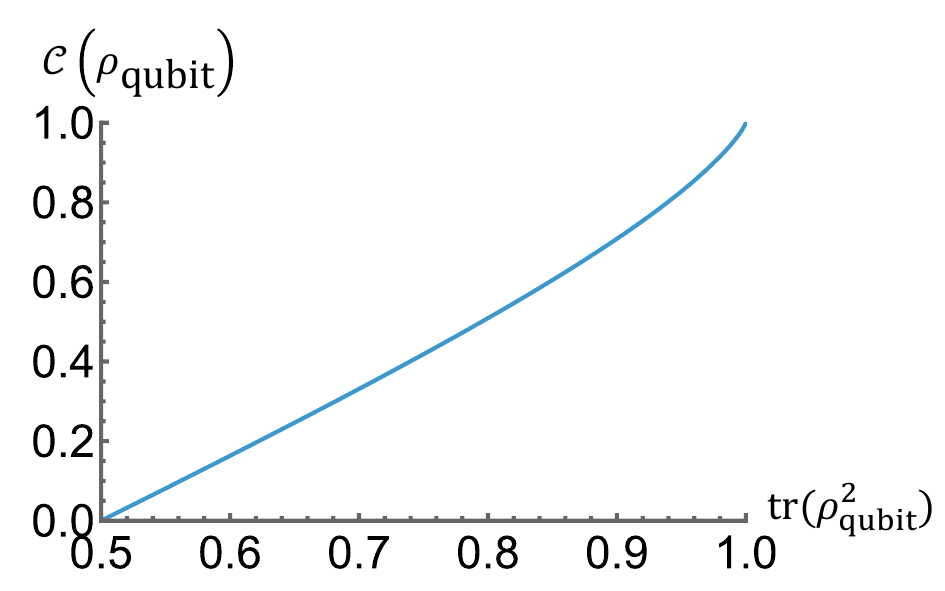}
	\caption{Complexity $\mathcal{C} (\rho_{\rm qubit})$ of a generic qubit state as a function of its purity ${\rm tr} (\rho_{\rm qubit}^2)$.}
	\label{fig:qubit}
\end{figure}

The simplest case $j=\frac12$ corresponds to a qubit system, where all pure states are coherent states, thus having a complexity of 1. Then we deal with mixed states. The general form of a (pure or mixed) qubit  state can be written as $\rho_{\rm qubit}=\frac{\mathbf{1}_2+\vec{r}\cdot \vec{\sigma}}{2}$, where $\vec{r}=(r_1,r_2,r_3)$ is a real three-dimensional vector such that $||\vec{r}||^2 =r_1^2+r_2^2+r_3^2\leq 1$, and $\vec{\sigma}=(\sigma_1,\sigma_2,\sigma_3)$ is the vector of the Pauli spin matrices \cite{nielsen2010quantum}. Due to rotational invariance, without loss of generality we can assume $\vec{r}=(0,0,r)$ with $0 \leq r \leq 1$, then the state becomes diagonal in the Dicke basis,
\begin{equation}
	\label{eq:rho_qubit}
	\rho_{\rm qubit} = \frac{1+r}{2} \left|\frac12,-\frac12\right\rangle \left\langle \frac12,-\frac12 \right| + \frac{1-r}{2} \left|\frac12,\frac12 \right\rangle \left\langle \frac12,\frac12 \right|,
\end{equation}
and its Husimi $Q$-function reads
\begin{equation*}
	Q(\Omega | \rho_{\rm qubit}) = \frac{1+r}{2} \cos^2 \frac{\theta}{2} + \frac{1-r}{2} \sin^2 \frac{\theta}{2}.
\end{equation*}
A straightforward calculation leads to the Wehrl entropy and Fisher information
\begin{eqnarray*}
	S_{\rm W} (\rho_{\rm qubit}) &=& \frac12 -\frac{1+r^2}{4r} \ln \frac{1+r}{1-r} -\frac12 \ln \frac{1-r^2}{4},\\
	I(\rho_{\rm qubit}) &=& 1- \frac{1-r^2}{2r}  \ln \frac{1+r}{1-r}.
\end{eqnarray*}
Thus the complexity of the state $\rho_{\rm qubit}$ is given by
\begin{equation}
    \label{eq:complexity_qubit}
	\mathcal{C} (\rho_{\rm qubit}) = e^{S_{\rm W} (\rho_{\rm qubit})-\frac12} I (\rho_{\rm qubit}) = \frac{2(\frac{1-r}{1+r})^{\frac{1+r^2}{4r}}}{\sqrt{1-r^2}} \left(1-\frac{1-r^2}{2r} \ln \frac{1+r}{1-r}\right).
\end{equation}
In Fig. \ref{fig:qubit}, we plot the complexity as a function of the purity ${\rm tr} (\rho_{\rm qubit}^2)=\frac{1+r^2}{2}$, from which it is clear that the complexity of a qubit state increases strictly in its purity.

\subsection{Dicke states}

Since the Dicke states $|j, \mu\rangle $ are pure, their amounts of complexity reduce to the entropy power $\mathcal{C} (j,\mu) = e^{S_{\rm W} (j,\mu) - \frac{2j}{2j+1}}$. From Eq. (\ref{eq:CSexpandDicke}) we immediately derive the Husimi function of a Dicke state as
\begin{equation}
	\label{eq:Husimi_Dicke}
    Q(\Omega | j,\mu)= {{2j} \choose {j-\mu}} \left( \cos^2 \frac{\theta}{2} \right)^{j-\mu} \left( \sin^2 \frac{\theta}{2} \right)^{j+\mu}.
\end{equation}
The Wehrl entropy can be calculated straightforwardly as
\begin{equation*}
    S_{\rm W} (j,\mu) = -\ln {{2j} \choose {j-\mu}} - (j-\mu) \psi(j-\mu+1) - (j+\mu) \psi(j+\mu+1) + 2j\psi(2j+2),
\end{equation*}
where $\psi(\cdot)$ is the digamma function. Note that it decreases in $|\mu|$. Therefore, among the Dicke states, the least complex ones are $|j,\pm j\rangle$ (in fact they are just coherent states, so $\mathcal{C} (j, \pm j)=1$), while the largest complexity is achieved by $|j,0\rangle$ in odd dimensions and $|j,\pm \frac12 \rangle$ in even dimensions.

\subsection{Thermal states}

Consider the thermal state
\begin{equation*}
	\rho_{\rm \beta} = \frac{1}{Z}  \sum_{\mu=-j}^j e^{-\beta \mu} |j,\mu\rangle \langle j,\mu|,
\end{equation*}
where $\beta=1/T$ denotes inverse temperature, and $Z=\sum_{\mu=-j}^j e^{-\beta \mu}$ is the partition function.
Its Wehrl entropy reads
\begin{equation*}
    S_{\rm W} (\rho_{\beta})= \ln \Big ( \frac{1-e^{-(2j+1)\beta}}{1-e^{-\beta}} \Big ) - 2j\beta\frac{e^{-(2j+1)\beta}}{1-e^{-(2j+1)\beta}} + \frac{2j}{2j+1},
\end{equation*}
and Fisher information reads
\begin{equation*}
    I(\rho_{\beta}) = \left\{ 
    \begin{array}{ll}
         2j- \frac{2j(2j+1)}{2j-1} e^{-\beta} \frac{1-e^{-(2j-1)\beta}}{1-e^{-(2j+1)\beta}}, & \quad j\neq\frac12, \\
         1- \frac{2\beta e^{-\beta}}{1-e^{-2\beta}}, & \quad j=\frac12.
    \end{array}
    \right.
\end{equation*}
Actually for $j=\frac12$ the thermal state is nothing but the state described by Eq. (\ref{eq:rho_qubit}), and for $j>\frac12$, we obtain the complexity of a thermal state as a function of its inverse temperature
\begin{equation}
	\mathcal{C} (\rho_{\beta}) = e^{-2j\beta \frac{e^{(2j+1)\beta}}{e^{(2j+1)\beta}-1}}\frac{(2j-1)(e^{(2j+1)\beta}-1)-(2j+1)e^\beta (e^{(2j-1)\beta}-1)}{(2j-1)(e^\beta-1)},
\end{equation}
which is an increasing function in $\beta$. Again, we would like to see the behavior against the purity of the thermal state
\begin{equation*}
	{\rm tr} (\rho_\beta^2) = \frac{(e^\beta -1)(e^{(2j+1)\beta}+1)}{(e^\beta +1)(e^{(2j+1)\beta}-1)}.
\end{equation*}

\begin{figure}[!h]
	\centering
	\includegraphics[width=0.5\textwidth]{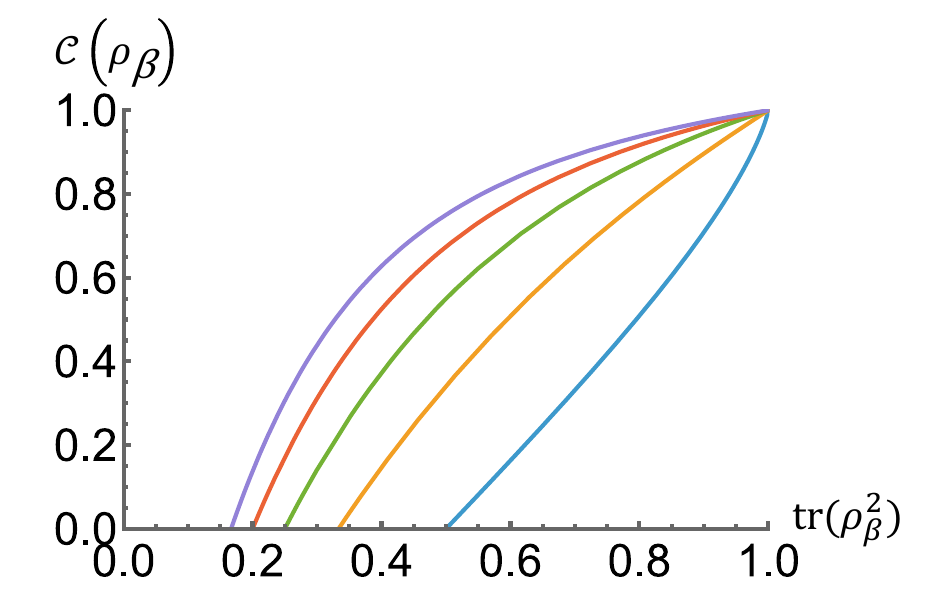}
	\caption{Complexity $\mathcal{C} (\rho_{\beta})$ of a Gibbs equilibrium state as a function of its purity ${\rm tr} (\rho_\beta^2)$ for $j=\frac12,1,\frac32,2,\frac52$ (from bottom to top). Note that the blue solid line for $j=\frac12$ coincides with Fig. \ref{fig:qubit} because a generic qubit state can be obtained from a Gibbs state via a rotation, which does not change complexity.}
	\label{fig:thermal}
\end{figure}

Fig. \ref{fig:thermal} shows that for each fixed dimension, the complexity of a thermal state is an increasing function in its purity. The lowest complexity is given by the completely mixed state $\rho_{\rm m}$ (i.e., $\beta=0$), while the supremum 1 is given by the coherent state $|j,-j\rangle$ as $\beta\to\infty$.

\subsection{Pure states with maximal Wehrl entropy}

\begin{figure}[!h]
	\centering
	\includegraphics[width=0.98\textwidth]{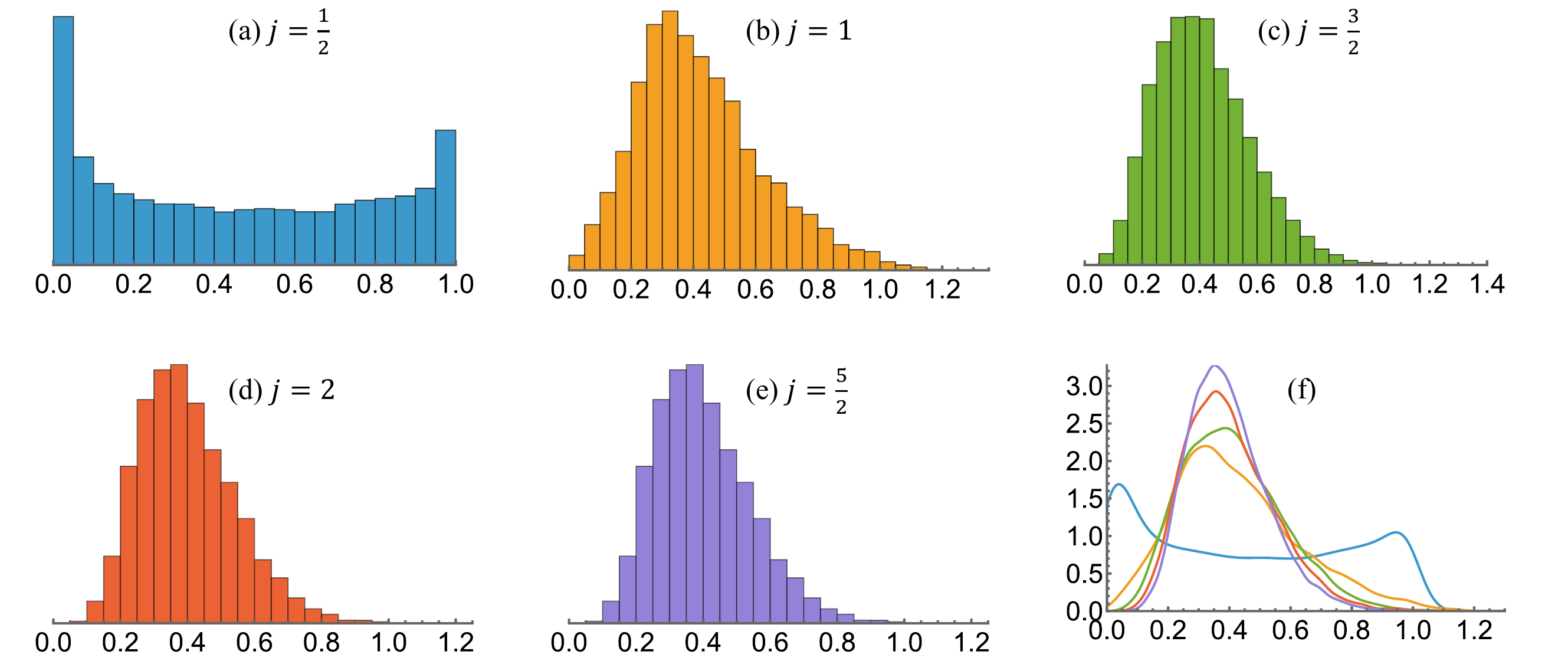}
	\caption{(a)-(e) Histograms of complexity. Each subfigure shows the complexity of $10^4$ randomly generated states for the given spin value $j$. (f) Smooth kernel histograms of (a)-(e), where the vertical axis shows the value of the probability density. When $j\geq1$, the most probable (typical) value of complexity is around 0.3 to 0.4, regardless of the spin value. }
	\label{fig:random_hist}
\end{figure}

In order to gain some insights on the typical values of complexity we generate $10^4$ random quantum states at each fixed dimension. In Fig. \ref{fig:random_hist} we present the histograms of the complexity of random states, from which we observe when $j\geq1$, a typical value of complexity occurs in the range [0.3,0.4], regardless of the dimension.

\begin{figure}[ht!]
	\centering
	\includegraphics[width=0.5\textwidth]{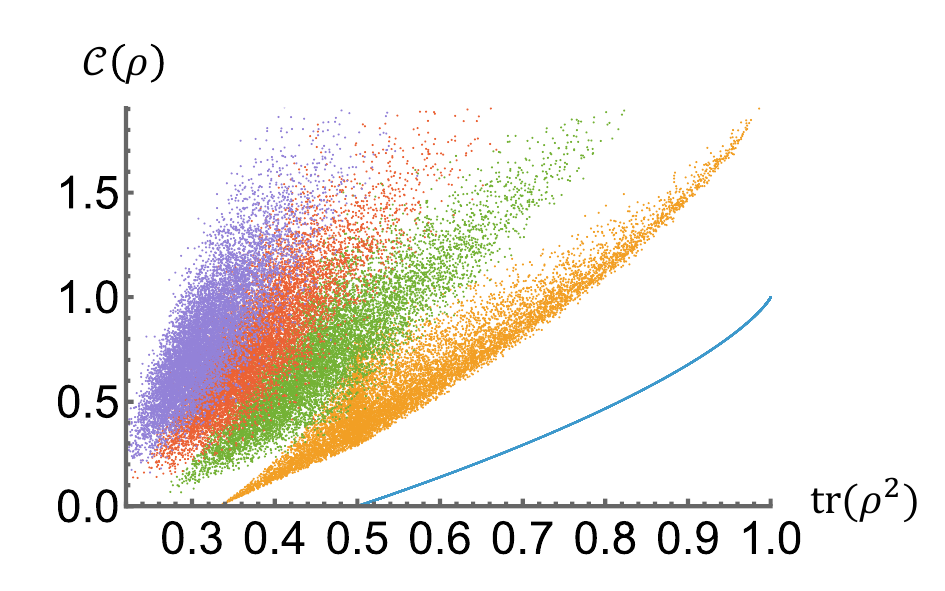}
	\caption{Complexity against purity. The plot reports the complexity of randomly generated states as a function of their purity. We have generated $10^4$ samples for each value $j$ of the spin. The blue solid line is for qubit ($j=\frac12$) and coincides with the blue solid line of Fig. \ref{fig:thermal}. The other points are for $j=1$ (orange), $j =\frac32$ (green), $j=2$ (red), $j=\frac52$ (purple). The complexity is fully determined by the purity only for qubits, whereas for higher dimensions purity is no longer the sole factor. Nevertheless, it is apparent from the plot the complexity is increasing with purity, suggesting the maximal complexity is achieved by pure states.}
	\label{fig:random_purity}
\end{figure}

In Fig. \ref{fig:random_purity} we plot the complexity of those random states against the purity. Note that only when $j=\frac12$, i.e., qubit systems, can the complexity be fully determined by the purity. For higher dimensions purity is no longer the sole factor. Nevertheless, it is clear that there is an increasing trend. Therefore we make the following conjecture.  

{\bfseries Conjecture 1.} The maximal complexity is achieved by pure states,
\begin{equation}
	\max_\rho \mathcal{C} (\rho) = \max_\psi \mathcal{C} (|\psi\rangle \langle \psi|).
\end{equation}
Since for pure states, the complexity reduces to the entropy power, finding the maximal complexity is equivalent to finding the maximal Wehrl entropy. This has been studied in Refs. \cite{baecklund2013maximization,Baecklund_2014,goldberg2020extremal} but remains an open problem. The idea is to consider the Majorana stellar representation \cite{majorana1932atomi}. Any pure state can be represented as $2j$ unordered points on the Bloch sphere. The coherent states are the only ones for which all points coincide, and they have the minimal Wehrl entropy \cite{lieb2014proof}. So naturally, the maximal Wehrl entropy is attained when the $2j$ points are spread far apart from each other \cite{lee1988wehrl,schupp1999lieb}. However, finding the optimal constellation is a difficult task in general. Below in Table \ref{tab:max_Wehrl} we list some numerical results for low dimensions. Note that the data about the maximal Wehrl entropy (second column) is taken from Ref. \cite{baecklund2013maximization}.

\begin{table}[!h]
	\centering
	\begin{tabular}{|c|c|c|}\hline
		$j$& $\max _{\psi} S_{\rm W} (\psi)$& $ \max _{\psi}\mathcal{C} (\psi)$\\\hline
		1&  0.973519& 1.3591\\\hline
		$\frac32$&  1.23871& 1.6302\\\hline
		2&  1.49166& 1.9970\\\hline
		$\frac52$& 1.65531& 2.2750\\\hline
		3&  1.83594& 2.6613\\\hline
		$\frac72$&  1.95286& 2.9384\\\hline
		4&  2.07789& 3.2838\\\hline
		$\frac92$&2.18494& 3.6145\\ \hline
	\end{tabular}
	\caption{Maximal Wehrl entropy $S_{\rm W}$ obtained by pure states $|\psi\rangle \in \mathcal{H}^j$ and the corresponding complexity $\mathcal{C} (\psi)=e^{S_{\rm W}(\psi)-2j/(2j+1)}$ for $j=1,\frac32,\cdots,\frac92$.}
	\label{tab:max_Wehrl}
\end{table}

\subsection{Spin squeezing}\label{sec:squeezing}

The concept of spin squeezing was discussed in Refs. \cite{kitagawa1993squeezed,wang2001spinsqueezing,wang2003spin,ma2011quantum}, with applications in quantum metrology \cite{wineland1992spin,wineland1994squeezed,gross2012spin}. The definition of spin squeezing is not unique. Here we follow Ref. \cite{kitagawa1993squeezed} and consider two types of nonlinear twisting Hamiltonians for generating spin squeezing: the one-axis twisting
\begin{equation*}
	H_1=\eta J_3^2,
\end{equation*}
with an initial state $|s_1(0)\rangle = |\theta=\frac{\pi}{2}, \phi=0 \rangle$, and the two-axis countertwisting
\begin{equation*}
	H_2= \eta(J_1J_2 +J_2J_1) =\frac{\eta}{2i} (J_+^2-J_-^2),
\end{equation*}
with an initial state $|s_2(0)\rangle = |\theta=0, \phi=0 \rangle =|j,-j\rangle$.
In the first case, the spin squeezed state $|s_1(\eta)\rangle$ can be expressed explicitly in the Dicke basis as
\begin{eqnarray*}
    |s_1(\eta)\rangle &:=& e^{-i \eta J_3^2} |\theta=\frac{\pi}{2}, \phi=0 \rangle\\
    &=& \frac{1}{2^j} \sum_{\mu=-j}^j \sqrt{{{2j} \choose {j-\mu}}} e^{-i \eta \mu^2 } |j,\mu\rangle.
\end{eqnarray*}
Then we can obtain the Husimi $Q$-function and numerically calculate the complexity of $|s_1(\eta)\rangle$. However for the countertwisting case it is difficult to derive a general expression for the expansion, and we can only provide the formula for some low dimensions.
\begin{eqnarray*}
    |s_2(\eta)\rangle &=& e^{- \eta (J_+^2 - J_-^2) /2} |j,-j\rangle\\
    &=& \left\{ 
    \begin{array}{ll}
         \cos \eta \left|1,-1 \right\rangle - \sin \eta \left|1,1 \right\rangle, & \quad j=1, \\
         \cos (\sqrt{3}\eta) \left|\frac32,-\frac32 \right\rangle - \sin (\sqrt{3}\eta) \left|\frac32,\frac12 \right\rangle, & \quad j=\frac32, \\
         \frac{1+\cos (2\sqrt{3}\eta)}{2} |2,-2\rangle -\frac{\sin (2\sqrt{3}\eta)}{\sqrt{2}} |2,0\rangle + \frac{1-\cos (2\sqrt{3}\eta)}{2} |2,2\rangle, & \quad j=2, \\
         \frac{9+5\cos (2\sqrt{7}\eta)}{14} \left|\frac52,-\frac52 \right\rangle - \frac{\sqrt{5}\sin (2\sqrt{7}\eta)}{\sqrt{14}} \left|\frac52,-\frac12 \right\rangle + \frac{\sqrt{45}(1- \cos (2\sqrt{7}\eta))}{14} \left|\frac52,\frac32 \right\rangle, & \quad j=\frac52, \\
         \cdots.
    \end{array}
    \right.
\end{eqnarray*}

\begin{figure}[!h]
	\centering
	\includegraphics[width=0.98\textwidth]{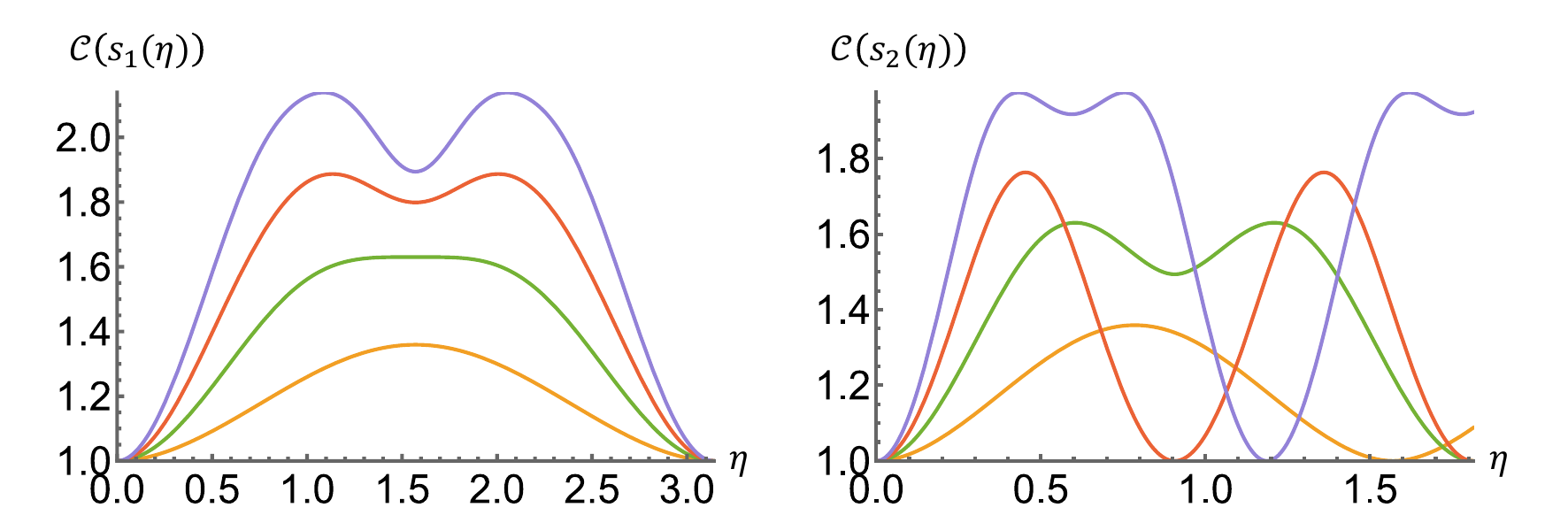}
	\caption{Complexity $\mathcal{C} (s_1(\eta))$ of one-axis spin-squeezed states (left) and $\mathcal{C} (s_2(\eta))$ of two-axis spin-squeezed states (right) as a function of the squeezing parameter $\eta$, for $j=1,\frac32,2,\frac52$ (from bottom to top).}
	\label{fig:squeezing}
\end{figure}

In Fig. \ref{fig:squeezing} we plot the complexity $\mathcal{C} (s_l(\eta))$, $l=1,2$, as a function of the squeezing parameter $\eta$ for $j=1,\frac32,2,\frac52$. Note that in qubit systems, all pure states are coherent states so squeezing is not possible.

We see that the complexity is a periodic function in $\eta$. For the one-axis-squeezed states, the period is $\pi$ for all dimensions, while for the two-axis-squeezed states the period depends on the dimension. Next, we look at the maximal amount of complexity generated by spin squeezing and we find that 
\begin{eqnarray*}
	\max_\eta \mathcal{C} (s_2(\eta)) &=& \max_\eta \mathcal{C} (s_1(\eta)) =\max _{\psi}\mathcal{C} (\psi),\qquad j=1,\frac32,\\
	\max_\eta \mathcal{C} (s_2(\eta)) &<& \max_\eta \mathcal{C} (s_1(\eta)) <\max  _{\psi}\mathcal{C} (\psi),\qquad j \geq 2,
\end{eqnarray*}
where $\max  _{\psi}\mathcal{C} (\psi)$ is the complexity of the optimal constellation given by Table \ref{tab:max_Wehrl}. This suggests that for $j=1,\frac32$, it is possible to achieve the maximal amount of complexity among pure states (or among all states if Conjecture 1 is true) by spin squeezing (either one- or two-axis-squeezing), while for larger $j$, spin squeezing is no longer an enough resource to generate the most complex states.

\subsection{NOON states}
The su(2) algebra has a bosonic representation, called the Schwinger realization. If one distributes a fixed number of $2j$ bosons over two modes with annihilation operators $a$ and $b$, respectively, the correspondence is made by \cite{sanders2014connection}
\begin{equation*}
	J_+=a^\dag b,\qquad J_-=b^\dag a, \qquad J_3=\frac12 (a^\dag a -b^\dag b).
\end{equation*}
The so-called NOON state is an entangled two-mode bosonic state defined as
\begin{equation*}
	|{\rm NOON}\rangle=\frac{1}{\sqrt{2}} (|N \rangle \otimes |0\rangle + |0 \rangle \otimes |N\rangle),
\end{equation*}
and is well-known as a resource for quantum metrology both theoretically \cite{bollinger1996optimal} and experimentally \cite{chalopin2018quantum}. Using the Schwinger realization a NOON state can be seen as a  superposition of SU(2) coherent states \cite{sanders2014connection},
\begin{equation}
	\label{eq:NOON_su2def}
	|{\rm NOON}\rangle=\frac{1}{\sqrt{2}} (|j,-j \rangle  + |j,j \rangle),
\end{equation}
with $j=N/2$, such that the Husimi $Q$-function can be immediately written  as
\begin{eqnarray*}
	Q(\Omega|{\rm NOON}) &=&\frac12 \left|\left( \cos \frac{\theta}{2} \right)^{2j} + \left( \sin \frac{\theta}{2} \right)^{2j} e^{2ij\phi} \right|^2 \\
	&=& \frac12 \left( \left( \cos \frac{\theta}{2} \right)^{4j} + \left( \sin \frac{\theta}{2} \right)^{4j} +2 \left( \cos \frac{\theta}{2} \right)^{2j} \left( \sin \frac{\theta}{2} \right)^{2j} \cos (2j\phi)\right).
\end{eqnarray*}
Again in this case the complexity only depends on the Wehrl entropy since the Fisher information is a constant. We compute the complexity for the NOON states in some low dimensions, and the results are reported in Table \ref{tab:NOON}. An interesting observation is that the NOON states can achieve the maximal complexity only when $j=\frac12,1,\frac32$, but not for higher dimensions. {This can be understood geometrically using the Majorana representation, as the NOON states are represented by $2j$ points distributed evenly on the equator (i.e., vertices of a regular $2j$-polygon) \cite{sanchezsoto2026quantumskymajoranastars}. Thus, for $j=\frac12,1,\frac32$, they are indeed the optimal constellation to separate the $2j$ points as far away from each other as possible (up to rotations). On the contrary, for higher dimensions we cannot achieve optimality by placing all the points on the same plane (for example, when $j=2$ the optimal constellation is given by a regular tetrahedron).} On the other hand, we note that
\begin{equation}
	\lim_{j \to \infty} \mathcal{C} ({\rm NOON}) =2.
\end{equation}
In fact, this is not completely surprising since the NOON states can be viewed as the DV analogue of the Schr\"odinger cat states in CV systems \cite{sanders2014connection}, and in Fig. 3 of Ref. \cite{Tang_2025} we see that the complexity of cat states also has a limit of 2 as the amplitude of the cats goes to infinity. The limit 2 comes from the fact that there are two components when expressing the state in terms of coherent states (Eq. (\ref{eq:NOON_su2def})), and that we have normalized the complexity of coherent states to be 1 in Definition 1.
\begin{table}[h!]
	\centering
	\begin{tabular}{|c|c|c|}\hline
		$j=N/2$& $S_{\rm W} ({\rm NOON})$& $\mathcal{C} ({\rm NOON})$\\\hline
		$\frac12$& 0.5& 1\\\hline
		1& 0.973519& 1.3591\\\hline
		$\frac32$& 1.23871& 1.6302\\\hline
		2& 1.38723& 1.7990\\\hline
		$\frac52$& 1.4722& 1.8943\\\hline
		3& 1.52266& 1.9455\\\hline
		$\frac72$&  1.55414& 1.9722\\\hline
		4& 1.57495& 1.9859\\\hline
		$\frac92$&  1.58957& 1.9929\\ \hline
	\end{tabular}
	\caption{Complexity of NOON states for different values of the spin.}
	\label{tab:NOON}
\end{table}

\section{Complexity of DV quantum channels}
In Ref. \cite{tang2026statistical} the complexity of CV quantum channels is defined as the ability to generate complexity from simple states such as coherent states. However in DV systems since coherent states are no longer the simplest states, after going through a quantum channel their complexity can either increase or decrease. Therefore we introduce two measures
\begin{eqnarray}
	\mathcal{C}_+ (\mathcal{E}) &=& \max \Big \{\sup_{\Omega \in S^2} \mathcal{C} (\mathcal{E} (|\Omega\rangle \langle \Omega |) ) -1 ,0 \Big \},\\
	\mathcal{C}_- (\mathcal{E}) &=& \max \Big \{1-\inf_{\Omega \in S^2} \mathcal{C} (\mathcal{E} (|\Omega\rangle \langle \Omega |) ) ,0 \Big\},
\end{eqnarray}
to capture the abilities of the quantum channel $\mathcal{E}$ to generate complexity and to destroy complexity, respectively.
First, we note that any unitary gate
\begin{equation*}
	\mathcal{E}_U (\rho) = U \rho U^\dag
\end{equation*}
will map coherent states to pure states, and the complexity of a pure state is at least 1, i.e., $\mathcal{C} (\psi) =e^{S_{\rm W}(\psi)-\frac{2j}{2j+1}} \geq 1$. This means that unitary gates cannot destroy complexity,
\begin{equation*}
	\mathcal{C}_- (\mathcal{E}_U) =0, \quad \forall \ U.
\end{equation*}
So we only need to consider their complexity generating power, $\mathcal{C}_+ (\mathcal{E}_U)$. But the case for qubit states ($j=\frac12$) is trivial: $\mathcal{C}_+ (\mathcal{E}_U) =0, \forall \ U$, because all pure states are coherent states, thus having a complexity of 1. Nontrivial examples occur when $j\geq 1 $. We consider some common unitary gates \cite{schwinger1960unitary}: the Pauli $X$ and $Z$ gates, the discrete Fourier transform $F$, the phase gate $P$,
\begin{eqnarray*}
	X&=& \sum_{\mu=-j}^{j-1} |j,\mu+1\rangle \langle j,\mu|+|j,-j\rangle\langle j,j|,\\
	Z&=& \sum_{\mu=-j}^j \left(e^{2\pi i / (2j+1)}\right)^{j+\mu} |j,\mu\rangle \langle j,\mu|,\\
	F&=& \frac{1}{\sqrt{2j+1}} \sum_{\mu',\mu=-j}^j \left(e^{2\pi i / (2j+1)}\right)^{(j+\mu')(j+\mu)}|j,\mu'\rangle \langle j,\mu| ,\\
	P&=& \sum_{\mu=-j}^j \left(-e^{\pi i/(2j+1)} \right)^{(j+\mu)^2} |j,\mu\rangle \langle j,\mu|,
\end{eqnarray*}
as well as the one-axis and two-axis squeezing operators we mentioned in Sec. \ref{sec:squeezing},
\begin{eqnarray*}
    S_1(\eta)&=& e^{-i\eta J_3^2},\\
    S_2(\eta)&=& e^{-\eta (J_+^2 - J_-^2)/2}.
\end{eqnarray*}
We calculate their complexity generating power $\mathcal{C}_+ (\mathcal{E}_\cdot)$ in Table \ref{tab:ugate}. Note that the Pauli $Z$ gate has zero complexity generating power because it is an SU(2) displacement operator, and according to Proposition 1, displacements will not change complexity. 
Furthermore, note that the phase gate $P$ is in fact a one-axis squeezing operator (up to an SU(2) displacement), so it is immediate that $\mathcal{C}_+ (\mathcal{E}_P) \leq \max_\eta \mathcal{C}_+ (\mathcal{E}_{S_1(\eta)})$.
The complexity generating power of the one-axis squeezing operators $S_1(\eta)$ is exactly as shown in Fig. \ref{fig:squeezing} (left), up to a constant of 1, and the initial coherent states which give the suprema are the ones on the equator ($\theta=0$). But for the two-axis squeezing operators  $S_2(\eta)$, maximizing over all coherent states could possibly lead to a higher complexity than that in Fig. \ref{fig:squeezing} (right).
\begin{table}[h!]
	\centering
	\begin{tabular}{|c|c|c|c|c|c|c|}\hline
		$j$& $\mathcal{C}_+ (\mathcal{E}_X)$& $\mathcal{C}_+ (\mathcal{E}_Z)$ &$\mathcal{C}_+ (\mathcal{E}_F)$ &$\mathcal{C}_+ (\mathcal{E}_P)$ &$\max_\eta \mathcal{C}_+ (\mathcal{E}_{S_1(\eta)})$
        &$\max_\eta\mathcal{C}_+ (\mathcal{E}_{S_2(\eta)})$\\\hline
		1& 0.3591& 0& 0.3214& 0.2741& 0.3591 & 0.3591\\\hline
		$\frac32$& 0.5854& 0& 0.6082& 0.4438& 0.6302 & 0.6302\\\hline
		2& 0.7467& 0& 0.7931& 0.5533& 0.8869 & 0.9380\\\hline
		$\frac52$& 0.8520& 0& 0.8945& 0.6214& 1.1389 &0.9748\\\hline
	\end{tabular}
	\caption{Complexity generating power $\mathcal{C}_+ (\mathcal{E}_U)$ of some unitary gates $U=X, Z, F, P, S_1(\eta), S_2(\eta)$.}
	\label{tab:ugate}
\end{table}

The next example is the amplitude damping channel, which describes the effects of energy dissipation \cite{nielsen2010quantum},
\begin{equation*}
	\mathcal{E}_p (\rho) = \sum_{n=0}^{2j} K_n \rho K_n^\dag, \qquad 0\leq p \leq 1,
\end{equation*}
where
\begin{eqnarray*}
	K_0&=& |j,-j\rangle \langle j,-j| + \sqrt{1-p} \sum_{\mu=-j+1}^{j} |j,\mu\rangle \langle j,\mu|, \\
	K_n&=& \sqrt{p} |j,-j\rangle \langle j,-j+n|,\qquad n=1,2,\cdots,2j.
\end{eqnarray*}

For qubits ($j=\frac12$), we can obtain analytic results since the complexity depends entirely on the purity, and the amplitude damping channel will reduce the purity of coherent states to
\begin{equation*}
	{\rm tr} \left( \mathcal{E}_p (|\Omega\rangle\langle\Omega|)^2 \right) = \frac14 \left(4-3p+3p^2+4p(1-p)\cos\theta -p(1-p)\cos (2\theta) \right),
\end{equation*}
which is a decreasing function in $\theta \in [0,\pi)$. So the maximal purity 1 is achieved when the initial state lies at the North pole $\theta=0$ (corresponding to the ground state $|j,-j\rangle$, which is left invariant by the amplitude damping channel), while the minimal purity is $p^2+(1-p)^2$, when the initial state lies at the South pole $\theta=\pi$ (corresponding to the excited state $|j,j\rangle$). Therefore the ability to destroy complexity is given by
\begin{equation*}
	\mathcal{C}_-(\mathcal{E}_p)=1-\mathcal{C} (\rho_{\rm qubit}),
\end{equation*}
where $\mathcal{C} (\rho_{\rm qubit})$ is given by Eq. (\ref{eq:complexity_qubit}), with $r=|1-2p|$. Then we see that the channel $\mathcal{E}_{1/2}$ has the greatest complexity-destroying power by turning the excited state to the completely mixed state $\rho_{\rm m}$.
On the other hand, the ability to generate complexity is always zero, i.e.,  $\mathcal{C}_+(\mathcal{E}_p)=0, \forall \ p$, since in a qubit system, all pure states are coherent states.

\begin{figure}[!h]
	\centering
	\includegraphics[width=0.98\textwidth]{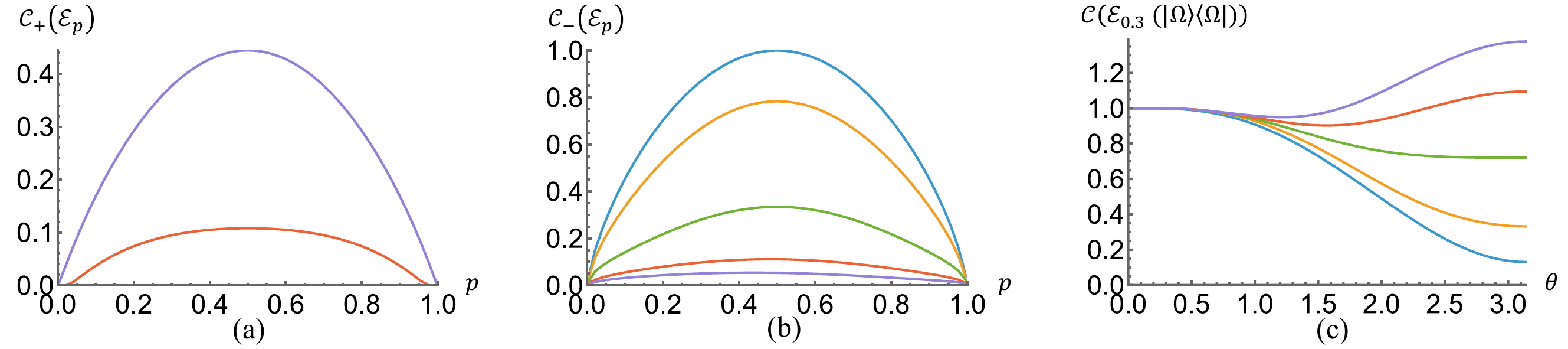}
	\caption{(a) Complexity generating power $\mathcal{C}_+(\mathcal{E}_p)$ of the amplitude damping channel $\mathcal{E}_p$ as a function of the damping probability $p$, for $j=2,\frac52$ (from bottom to top). For $j=\frac12,1,\frac32$, the curves vanished, i.e., $\mathcal{C}_+(\mathcal{E}_p)\equiv0$. (b) Complexity breaking power $\mathcal{C}_-(\mathcal{E}_p)$ of the amplitude damping channel $\mathcal{E}_p$ as a function of the damping probability $p$, for $j=\frac12,1,\frac32,2,\frac52$ (from top to bottom). (c) Complexity $\mathcal{C} (\mathcal{E}_{0.3}(|\Omega\rangle\langle\Omega|))$ of a coherent state $|\Omega=(\theta,\phi)\rangle$ after going through the amplitude damping channel $\mathcal{E}_{0.3}$, as a function of its polar angle $\theta$, for $j=\frac12,1,\frac32,2,\frac52$ (from bottom to top). Note that when $j\geq2$, the curves are not monotonically decreasing in $\theta$.}
	\label{fig:channelAD}
\end{figure}

For higher dimensions, we plot the two measures, $\mathcal{C}_+(\mathcal{E}_p)$ and $\mathcal{C}_-(\mathcal{E}_p)$, as a function of $p$ in Fig. \ref{fig:channelAD}(a)-(b). We see that when the damping probability $p$ is 1/2, the channel becomes the most versatile in both generating and destroying complexity, depending on the input state. Besides, from its definition we always have $\mathcal{C}_- (\cdot) \leq 1$, and we see that only in the case of qubit systems can the amplitude damping channel achieve this upper bound: $\mathcal{E}_{1/2}$ can completely destroy the complexity. While in higher dimensions, $\mathcal{E}_{1/2}$ may only output the mixture $\frac12|j,-j\rangle\langle j,-j|+\frac12|j,j\rangle\langle j,j|$, but not the completely mixed state $\rho_{\rm m}$, so it is not able to destroy all the complexity. The most surprising observation is that although the amplitude damping channels have no complexity-generating power in low dimensions $j=\frac12,1,\frac32$, they do in higher dimensions. In fact, if we fix some $p\in [0,1]$ and look at the complexity of the output state as a function of the polar angle $\theta$ of the input coherent state (we note that the complexity does not depend on the azimuthal angle $\phi$) in Fig. \ref{fig:channelAD}(c), we see that for $j\geq 2$, the trend is no longer monotonically decreasing, so it is possible that some coherent states will become more complex after going through the amplitude damping channel. On the contrary, the purity of the output state is always monotonically decreasing in $\theta$ in any dimension. This further shows that in high dimensions, complexity depends on other factors than purity. 

\section{Conclusions}
In this work, we have extended the phase-space approach to statistical complexity from continuous-variable systems to the discrete-variable regime. By constructing a Husimi $Q$-function over spin coherent states on the Bloch sphere, we have defined a complexity quantifier of quantum states capturing the trade-off between localization and delocalization in SU(2) systems. The measure combines the Wehrl entropy—which quantifies phase-space spread—and the Fisher information—which quantifies sensitivity to displacements—into a single scalar. 

Our analysis reveals several qualitative differences between DV and CV complexity. Unlike in the CV case, where coherent states are the minimizers of complexity, in DV systems the completely mixed state attains zero complexity while coherent states serve as a fixed unit of reference. This distinction is ultimately geometric: the compactness of SU(2) guarantees finite-dimensional representations and normalizable uniform distributions, whereas the non-compact nature of the Weyl–Heisenberg group precludes such a state in CV systems. Conversely, while CV complexity is unbounded above, DV complexity is bounded for each fixed dimension, with the supremum conjectured to be attained by pure states.

We have provided a comprehensive analysis of the complexity across a wide range of physically relevant states. These results support the conjecture that maximal complexity is achieved by pure states, thereby reducing the problem to the optimization of Wehrl entropy over Majorana constellations (i.e., finding an optimal constellation for $2j$ unordered points on the Bloch sphere being spread as far apart from each other as possible), which is a long-standing open problem in quantum optics and geometric quantum mechanics, and also is intimately related to spherical design and distributing points uniformly on spheres \cite{cui1997equidistribution, lubotzky1986hecke, seymour1984averaging, saff1997distributing, katanforoush2003distributing}. We have further shown that spin squeezing and NOON states, despite being paradigmatic resources for quantum metrology, can saturate this maximal complexity only in low dimensions. For $j\geq2$, they fall short, indicating that the generation of maximal complexity requires more sophisticated state preparations that cannot be realized by these mechanisms alone.

We have also extended the framework to quantum channels, introducing separate measures for the ability to generate and to destroy (break) complexity. For unitary gates, we find that complexity generation is nontrivial only for $j\geq 1$, while complexity destruction is identically zero. For the amplitude damping channel, we uncover a rich phenomenology: in low dimensions, the channel can only destroy complexity; in higher dimensions, it can also generate complexity from coherent inputs. This behavior is not captured by purity alone and underscores the need for phase-space based diagnostics beyond simple mixedness measures.

Several open questions remain. First, while our conjecture that maximal complexity is attained by pure states is strongly supported by numerical evidence, a rigorous proof, or a counterexample, is highly desirable. Second, the precise relationship between complexity and other measures of quantumness is not yet fully understood and warrants further investigation. Third, the complexity of quantum channels introduced here may be extended to incorporate dynamical complexity, and operational interpretations in terms of resource theories. Finally, the connection between optimal Majorana constellations and complexity suggests a deeper link between phase-space geometry and information theory that may extend to other Lie groups and homogeneous spaces.

In summary, we have established a unified framework for quantifying phase-space complexity that now spans both continuous and discrete variable systems. By revealing the dimension-dependent limitations of common quantum resources and the subtle interplay between purity, coherence, and phase-space structure, our results contribute to a broader understanding of what makes a quantum state, or a quantum process, statistically complex.

\acknowledgments
This work was supported by Beijing Natural Science Foundation, Grant No. Z250004, the National Natural Science Foundation of China, Grant No. 12341103. Support from the NFPO organization {\em Comitato Quantum} is also acknowledged.

\appendix

\section{Fisher information for pure states}
Here we prove that $I(\psi)=2j$ for any pure state $|\psi\rangle \in \mathcal{H}^j$. The proof follows a similar argument as in Ref. \cite{luo2001fisher}.
Recall that a pure state has a Bargmann representation in the space $\mathcal{F}^j$ of the polynomials of degree less than $2j+1$, i.e.,
\begin{equation*}
    \mathcal{F}^j = \{f(z)=a_0+a_1z+\cdots+a_{2j}z^{2j}:a_0,a_1,\cdots,a_{2j}\in\mathbb{C}\},
\end{equation*}
where the inner product in $\mathcal{F}^j$ is defined as \cite{Perelomov1977}
\begin{equation}
    \label{eq:bargmann_norm}
    \langle f_1, f_2\rangle = \int _{\mathbb{C}}\overline{f_1(z)} f_2(z) \frac{1}{(1+|z|^2)^{2j+2}} \frac{2j+1}{\pi} {\rm d}^2 z,
\end{equation}
where ${\rm d}^2z={\rm d}x{\rm d}y$ for $z=x+iy\in \mathbb{C}, x,y\in \mathbb{R}.$
In fact, for any $|\psi\rangle \in \mathcal{H}^j$, its Bargmann representation can be obtained via
\begin{equation*}
    |\psi\rangle \mapsto f(z)=(1+|z|^2)^j \langle \bar{z} |\psi\rangle \in \mathcal{F}^j,
\end{equation*}
where $|z\rangle$ denotes the coherent state $|\Omega\rangle$ with stereographic projection
\begin{equation*}
    z=\tan \frac{\theta}{2} e^{i \phi}.
\end{equation*}
Then, the Husimi $Q$-function for the pure state $|\psi\rangle$ reads
\begin{equation}
    Q(\Omega|\psi)= |\langle \Omega|\psi\rangle|^2 = (1+|z|^2)^{-2j} |f(z)|^2.
\end{equation}
On the other hand, note that the Fisher information defined in Eq. (\ref{eq:def_Fisher}) can be written equivalently as
\begin{equation*}
    I(\rho) = \int _{S^2}4 \Big|\Big|\nabla \sqrt{Q(\Omega | \rho)} \Big|\Big|^2 \frac{2j+1}{4\pi} {\rm d}\Omega,
\end{equation*}
so we calculate the partial derivatives of $\sqrt{Q}=(f\bar{f})^{\frac12}(1+|z|^2)^{-j}$ and get
\begin{eqnarray*}
    \partial_\theta \sqrt{Q} &=& \frac12 (f\bar{f})^{-\frac12} (f \cdot \partial_\theta\bar{f}+ \partial_\theta f \cdot \bar{f})(1+|z|^2)^{-j} -j (f\bar{f})^{\frac12}(1+|z|^2)^{-j} \tan \frac{\theta}{2} \\
    &=& \frac12 (f\bar{f})^{-\frac12} \left(f\cdot \frac12 e^{-i\phi} \Big (1+\tan^2 \frac{\theta}{2}\Big ) \bar{f'}  +  \frac12 e^{i\phi} \Big (1+\tan^2 \frac{\theta}{2}\Big ) f' \cdot \bar{f} \right)\\
    & &\cdot (1+|z|^2)^{-j} -j (f\bar{f})^{\frac12}(1+|z|^2)^{-j} \tan \frac{\theta}{2},\\
    \partial_\phi \sqrt{Q} &=& \frac12 (f\bar{f})^{-\frac12} (f \cdot \partial_\phi\bar{f}+ \partial_\phi f \cdot \bar{f})(1+|z|^2)^{-j} \\
    &=& \frac12 (f\bar{f})^{-\frac12} \Big (-f \cdot i \tan \frac{\theta}{2} e^{-i \phi}\bar{f'}+ i\tan \frac{\theta}{2} e^{i \phi} f' \cdot \bar{f} \Big ) (1+|z|^2)^{-j}.
\end{eqnarray*}
Then after some straightforward simplifications, we have
\begin{eqnarray*}
    ||\nabla \sqrt{Q}||^2 &=& \left(\partial_\theta \sqrt{Q}\right)^2+\frac{1}{\sin^2 \theta} \left(\partial_\phi \sqrt{Q}\right)^2 \\
    &=& j^2 f\bar{f} |z|^2  (1+|z|^2)^{-2j} -\frac{j}{2} (f\bar{f'}\bar{z} +f'\bar{f}z) (1+|z|^2)^{-(2j-1)} +\frac14 f'\bar{f'} (1+|z|^2)^{-(2j-2)},
\end{eqnarray*}
and hence the Fisher information is
\begin{eqnarray*}
    I(\psi) &=& \int _{S^2} 4 ||\nabla \sqrt{Q} ||^2 \frac{2j+1}{4\pi} {\rm d}\Omega \\
    &=& \int _{\mathbb{C}} 4 ||\nabla \sqrt{Q} ||^2 \frac{2j+1}{\pi} \frac{1}{(1+|z|^2)^2} {\rm d}^2z \\
    &=& \int _{\mathbb{C}}\left( 4j^2 f\bar{f} z\bar{z} -2j(f\bar{f'}\bar{z} +f'\bar{f}z) (1+z\bar{z}) + f' \bar{f'} (1+z\bar{z})^2 \right)  \frac{2j+1}{\pi} \frac{1}{(1+|z|^2)^{2j+2}} {\rm d}^2z \\
    &=& 4j^2 \langle zf, zf\rangle -2j \left( \langle zf',f\rangle +\langle f,zf'\rangle + \langle z^2f', zf \rangle + \langle zf, z^2 f'\rangle \right)\\
    & & + \langle f', f'\rangle +2 \langle zf',zf' \rangle +\langle z^2f',z^2f'\rangle.
\end{eqnarray*}
Note that in $\mathcal{F}^j$, the infinitesimal operators $J_3, J_{\pm}$ are represented by the following first-order differential operators \cite{Perelomov1977}
\begin{equation*}
    J_3=z \frac{{\rm d}}{{\rm d}z}-j, \qquad
    J_+=-z^2 \frac{{\rm d}}{{\rm d}z} +2jz, \qquad
    J_-=\frac{{\rm d}}{{\rm d}z}.
\end{equation*}
Furthermore, $J_+$ and $J_-$ are conjugated with respect to the norm Eq. (\ref{eq:bargmann_norm}), and $J_3$ is self-conjugated. Also note that $\frac{J_+J_-+J_-J_+}{2}+J_3^2=J_1^2+J_2^2+J_3^2=j(j+1)\mathbf{1}_{2j+1}$, and that $f$ is a unit vector, i.e., $\langle f,f \rangle=1$. Finally, using these relations, the last line in the above equation can be simplified to
\begin{equation}
    I(\psi)=2j.
\end{equation}

\section*{References}



\bibliography{ref}

\end{document}